\documentclass[aps,pra,twocolumn,halfspacing,10pt,nofootinbib]{revtex4-2}

\usepackage{epsfig}
\usepackage{dcolumn}
\usepackage{natbib}
\usepackage[utf8]{inputenc}
\usepackage[english]{babel}
\usepackage{xcolor}
\usepackage{graphicx}
\usepackage{color}
\usepackage{bm}
\usepackage{amssymb}
\usepackage[intlimits]{amsmath}
\usepackage{booktabs}
\usepackage{amsmath}
\usepackage[para,online,flushleft]{threeparttable}

\usepackage[colorlinks=true,allcolors=blue]{hyperref}
\usepackage{physics}
\usepackage{multirow}

\begin{document}
\title{Enhanced Schiff and magnetic quadrupole moments in deformed nuclei 
%Nuclei with an Octupole Deformation 
and their connection to the search for axion dark matter}
\author{F. Dalton$^{1}$}
\email{f.dalton@student.unsw.edu.au}
\author{V. V. Flambaum$^{1}$} 
\email{v.flambaum@unsw.edu.au}
\author{A. J. Mansour$^1$}
\email{andrew.mansour@student.unsw.edu.au}
\affiliation{$^1$School of Physics, University of New South Wales,
Sydney 2052, Australia}

\begin{abstract}
Deformed nuclei possess enhanced moments violating time reversal invariance ($T$) and parity ($P$). Collective magnetic quadrupole moments (MQM) appear in nuclei with a quadrupole deformation (which have ordinary $T$,$P$-conserving collective electric quadrupole moments).   Nuclei with an octupole deformation have a collective electric octupole moment, electric dipole moment (EDM), Schiff moment and MQM in the intrinsic frame which rotates with the nucleus. In a state with definite angular momentum in the laboratory frame, these moments are forbidden by $T$ and $P$ conservation, meaning their expectation values vanish due to nuclear rotation. However, nuclei with an octupole deformation have doublets of close opposite parity rotational states with the same spin, which are mixed by $T$,$P$-violating nuclear forces. This mixing polarises the orientation of the nuclear axis along the nuclear spin, and all moments existing in the intrinsic frame appear in the laboratory frame (provided the nuclear spin $I$ is sufficiently large  to allow such a moment). Such a mechanism produces enhanced $T$,$P$-violating nuclear moments. This enhancement also takes place in nuclei with a soft octupole vibration mode. In this paper we present updated estimates for the enhanced Schiff moment in isotopes of Eu, Sm, Gd, Dy, Er, Fr, Rn, Ac, Ra, Th, Pa, U, Np and Pu in terms of the CP-violating $\pi$-meson- nucleon interaction constants $\bar{g}_{0},\bar{g}_{1}$ and $\bar{g}_{2}$, the QCD parameter $\bar{\theta}$ and the quark chromo-EDMs. The implications of the enhanced $T$,$P$-violating moments to the search for axion dark matter in solid state experiments are also discussed, with potential alternative candidate compounds in which we may expect enhanced effects suggested. 

% Our results may be used to test $CP$-violation theories and may have applications in the search for axion dark matter in atomic, molecular and solid state experiments

\end{abstract}
\date{\today}
\maketitle

\section{Introduction}
Measurements of atomic and molecular time-reversal ($T$) and parity ($P$) violating electric dipole moments (EDMs) are used to test unification theories predicting the violation of fundamental symmetries $CP$ (invariance under combined operations of spatial inversion and charge reversal).  Such measurements have been able to exclude a number of models and significantly reduce the parametric space of other popular models including supersymmetry~\cite{PR,ERK}. This research also may shed light on the baryogenesis problem, the matter-antimatter asymmetry in the universe produced by an unknown $CP$-violating interaction. The expected magnitude of an EDM is small, so it is favourable to use mechanisms which enhance the effect, see e.g.~\cite{Khriplovich1991,Ginges2004,Khriplovich1997}. 

The nuclear EDM is completely screened in neutral atoms and molecules. However, a non-zero atomic EDM due to the nucleon EDM may still be produced if the distribution of the nucleon EDM %(directed along the axis of nucleon spin)
and charge in a nucleus are not proportional to each other~\cite{Schiff1963}. Nuclear EDMs may also be produced by $T$,$P$-violating nuclear forces, however in this case the mechanism is different as the resultant EDM is from the charge distribution in the entire nucleus~\cite{Sushkov1984}. The post-screening nucleus-electron interaction is proportional to the so called {\it Schiff moment}, a vector multipole which produces an electric field inside the nucleus, see~\cite{Sandars1967,Hinds1980,Sushkov1984,Sushkov1985,Sushkov1986}. The finite nuclear size corrections and distribution of the Schiff moment induced electric field inside the nucleus have been  found in Ref.~\cite{Ginges2002}. This resultant electric field polarises the atom and produces an atomic EDM in the direction of the nuclear spin $I$. It was shown in Refs.~\cite{Sushkov1984,Sushkov1985,Sushkov1986} that the leading contribution to the nuclear EDM and Schiff moment is the result of $T$,$P$-violating nuclear forces. 

In nuclei with spin $I \geq 1$, there exists another contribution to this atomic EDM, which yields from the nuclear nuclear magnetic quadrupole moment (MQM). The magnetic interaction between the MQM and atomic electrons mixes electron orbitals of opposite parity and produces an atomic EDM and T,P-violating nuclear spin - molecular axis interaction constants for molecules in electronic states with non-zero electron angular momentum \cite{Sushkov1984}. This magnetic interaction is not screened, so generically (without special enhancement factors) the atomic EDM produced by the interaction between nuclear MQM and electrons is expected to be an order of magnitude bigger than the EDM produced by the Schiff moment \cite{Sushkov1984} and two orders of magnitude bigger than the EDM produced by the electric octupole moment \cite{Murray1997}.

% For the nuclear EDM produced by $T,P$- violating nuclear forces the mechanism is different as this is the EDM of the charge distribution in the whole nucleus \cite{Sushkov1984}.
%However, a non-zero atomic EDM may still be produced if the distribution of the EDM and charge in a nucleus are not proportional to each other~\cite{Schiff1963}. 

Given the relatively small magnitude of the resultant atomic EDM, it is advantageous to seek the existence of a mechanism which enhances these nuclear moments. In this paper, we will focus on mechanisms of enhancement which arise via the study of deformed nuclei. A specific mechanism of enhancement of the Schiff moment may be found through the study of nuclei with an octupole deformation. Such nuclei exhibit a low-lying energy doublet to the ground state, with opposing parity and identical spin. The existence of such a state may lead to an enhancement of the nuclear Schiff moment produced by $T$,$P$-violating nuclear forces, which admix these two states. This mechanism may also be applicable to nuclei with a soft octupole vibration mode, however we should expect to see the largest enhancement ($\sim 10^{2} - 10^{3}$ times) in nuclei with an intrinsic octupole deformation, see~\cite{Auerbach1996,Spevak1997}. According to Refs.~\cite{Auerbach1996,Spevak1997} this occurs in specific isotopes of Fr, Rn, Ra and other actinide atoms. Atomic and molecular EDMs produced by Schiff moments increase with the nuclear charge $Z$ faster than $Z^{2}$~\cite{Sushkov1984}. This is another contributing factor to the expectation that EDMs in actinide atoms and their molecules are significantly larger than in other systems. 

Note that the enhancement of T,P-violating effects due to octupole doublets has been noted in Ref.\cite{Spevak1995}. Earlier such enhancement of P-odd effects was noted in Ref.~\cite{Sushkov1980}.

The Schiff moment is proportional to the product of the quadrupole deformation parameter and square of the octupole deformation parameter, $ \beta_{2} \beta_{3}^{2}$ \cite{Spevak1997}.  A comprehensive tabulation of the parameters $\beta_{2}$ and $\beta_{3}$ for the ground states of nearly all nuclei can be found in Ref.~\cite{Moller2016}. Estimates of these parameters with alternative nuclear structure models can give differing results, see e.g.~\cite{Ebata2017,Nazarewicz1984}.

Ref.~\cite{Spevak1997} estimates the squared value of the static octupole deformation parameter to be about $\beta_{3}^{2} \sim (0.1)^{2}$. According to Ref.~\cite{Engel2000}, in nuclei with a soft octupole vibration mode, the squared dynamical octupole deformation parameter also takes similar values $\ev{\beta_{3}^{2}} \sim (0.1)^{2}$. As such, a similar enhancement of the Schiff moment also may occur due to a dynamical octupole vibration mode~\cite{Engel2000,Zelevinsky2003,Auerbach2006} in nuclei with $\ev{\beta_{3}} = 0$~\footnote{Consider, for example, the displacement of an ordinary harmonic oscillator. In such a system, $\ev{x} = 0$, while $\ev{x^{2}} \neq 0$.}. This observation significantly increases the list of nuclei in which the Schiff moment is enhanced. 

Refs.~\cite{Auerbach1996,Spevak1997,Engel2003,Dobaczewski2018} have performed numerical calculations of the Schiff moments and estimates of the atomic EDM produced by the electrostatic interaction between electrons and these moments for $^{221}$Fr, $^{223}$Fr, $^{223}$Ra, $^{223}$Rn, $^{225}$Ra, $^{225}$Ac and  $^{229}$Pa. The short lifetime of these nuclei proves to be an issue when it comes to experiments. Such experiments for $^{223}$Rn and $^{225}$Ra have been considered by various experimental groups~\cite{Parker2015,Bishof2016,Tardiff2008}. Despite the Schiff moment enhancement, the $^{225}$Ra EDM measurements presented in Refs.~\cite{Bishof2016,Tardiff2008} have not reached the sensitivity to the $T$,$P$-violating interaction as the Hg EDM experiment~\cite{Graner2016}. These experiments are also hindered by the instability of $^{225}$Ra (which has a half-life of 15 days) and a relatively small number of atoms. Ref.~\cite{Feldmeier2020} extended the list of candidate nuclei to include stable isotopes $^{153}$Eu, $^{155}$Gd, $^{161}$Dy and $^{163}$Dy, as well as long lifetime nuclei $^{153}$Sm,$^{165}$Er,$^{225}$Ac, $^{227}$Ac, $^{229}$Th, $^{229}$Pa, $^{233}$U, $^{235}$U, $^{237}$Np and $^{239}$Pu.

% IMPORTANT
%-------
% The nuclear EDM, Schiff, electric octupole and magnetic quadrupole moments produced by T,P-odd nuclear forces are enhanced due to an opposite parity level with the same spin close to the ground state \cite{Haxton1983,Sushkov1984,Murray1997}. The collective enhancement of magnetic quadrupole moments in nuclei with a quadrupole deformation has been demonstrated in \cite{Flambaum1994} (see also \cite{Demille2014,Lackenby2018}). Calculations of the enhanced MQM in octupole deformed nuclei with spin $I \geq 1$ were completed in Ref.~\cite{Mansour2022}.

We may also seek a mechanism which enhances the contribution of the nuclear MQM to the atomic EDM. The enhancement of this effect in nuclei with a quadrupole deformation has been demonstrated in \cite{Flambaum1994}, in which it is shown that the $T$,$P$-violating interaction produces a collective MQM, inducing $T$,$P$-violating effects in atoms and molecules. Estimates of the enhanced effect in paramagnetic molecules TaN, ThO, ThF$^{+}$, HfF$^{+}$, YbF, HgF, and BaF are presented in Ref.~\cite{Demille2014}.  Accurate molecular calculations were performed for molecules of interest: ThO~\cite{Skripnikov2014}, TaN~\cite{Skripnikov2015}, TaO$^{+}$~\cite{Fleig2017}, HfF$^{+}$~\cite{Petrov2018}, YbOH~\cite{Maison2019}  and LuOH$^{+}$~\cite{Maison2020}. Further, Ref.~\cite{Lackenby2018} calculated the enhanced collective MQM in the quadrupole deformed nuclei $^{9}$Be, $^{21}$Ne, $^{27}$Al, $^{151}$Eu, $^{153}$Eu, $^{163}$Dy, $^{167}$Er, $^{173}$Yb, $^{177}$Hf, $^{179}$Hf, $^{181}$Ta, $^{229}$Th, and the resultant MQM induced energy shift in various diatomic molecules of experimental interest which contain these nuclei. An additional enhanced contribution to the MQM appears in octupole deformed nuclei via the aforementioned octupole mechanism. Calculations of the enhanced MQM in this mechanism are presented in Ref.~\cite{Mansour2022}.

There are various ways in which one may probe $T,P$-violating atomic, molecular and neutron EDMs. For example, one may seek to detect these effects by measuring the precession of the angular momentum of a system in the presence of an external electric field, analogous to Larmor precession in an external magnetic field. Another viable method of detection, first studied in Refs.~\cite{Shapiro1968,Kolycheva1978} involves the use of a condensed matter sample with $N$ spins in an external electric field. In this mechanism, any interaction of the field with the EDM's in the sample will lead to a slight alignment of the spins in the direction of the field, which may be in turn measured via probing the induced magnetisation of the sample. Such experiments have been the subject of a renewed interest~\cite{Lamoreaux2002,Liu2004,Mukhamedjanov2005}, with various experiments conducted constraining the value of the electron EDM~\cite{Heidenreich2005,Kim2011,Eckel2012,Kim2015}.

This method may also be employed in the search for nuclear Schiff moments. In particular, Ref.~\cite{Mukhamedjanov2005} proposed using the latter method with the ferroelectric sample PbTiO$_{3}$. Although using this sample would not benefit from any enhancement due to the octupole mechanism, an enhancement of the effect is expected due to the strong internal electric field of the ferroelectric, as well as the ability to cool the nuclear spin subsystem in this compound down to nanokelvin temperatures, which may limit the effect of phonons and suppress the nuclear spin-lattice interaction. Further, a study of the nuclear spin relaxation in $^{207}$Pb conducted in Ref.~\cite{Bouchard2008} reported a dramatic increase in the spin relaxation time as the temperature of the system is lowered, which may prove to be even further advantageous in the search for nuclear Schiff moments using the experiment proposed in Ref.~\cite{Mukhamedjanov2005}.  

Any enhancement of the aforementioned effects also has implications in the search for ultralight dark matter. The $CP$ violating neutron EDM may be due to the QCD $\theta$-term. It was noted in Ref.~\cite{Graham2011} that axion dark matter produces an oscillating neutron EDM, as the axion field is equivalent to an oscillating ${\bar \theta}$. The QCD $\theta$-term also produces $T$,$P$-violating nuclear forces, creating $T$,$P$-violating nuclear moments. Correspondingly, the axion field also produces oscillating nuclear $T$,$P$-violating moments~\cite{Stadnik2014}. To obtain results for the oscillating $T$,$P$-violating moments it is sufficient to replace the constant ${\bar \theta}$ by ${\bar \theta}(t)= a(t)/f_a$, where $f_a$ is the axion decay constant, $a(t) =a_0 \cos{m_a t}$, $(a_0)^2= 2 \rho /(m_a)^2$, where $\rho$ is the axion dark matter energy  density \cite{Graham2011,Stadnik2014}.  Moreover, in the case of a resonance between the frequency of the axion field oscillations and molecular transition frequency there may be a resonance enhancement of the oscillating nuclear $T$,$P$-violating moment effect~\cite{OscillatingEDM}.  As oscillating nuclear $T$,$P$-violating moments may be produced by axion dark matter, corresponding measurements may be used to search for this dark matter. This research is in progress, and the first results have been published in Ref.~\cite{nEDM}, in which the oscillating neutron EDM and oscillating $^{199}$Hg Schiff moment have been measured.  The effect produced by the oscillating axion-induced Pb Schiff moment in solid state materials has been measured by the Cosmic Axion Spin Precession Experiment (CASPEr) collaboration in Ref.~\cite{CasperNew}. The effect of oscillating $T$,$P$-violating nuclear polarisability  has been measured in Ref.~\cite{Cornell2021} (see theory in Refs.~\cite{pol1,pol2,pol3}). Further, oscillating MQMs were shown to produce resonance transitions in molecules in Ref.~\cite{Budker2020}.

% We note that this mechanism of enhancement may also be extended to other nuclear moments. 

This paper is organised as follows. In Section \ref{EstimatesSection}, we perform estimates of the nuclear Schiff moment in various stable and long-living nuclei theorised to have either a static octupole deformation or a dynamical octupole vibration mode, using values of $\beta_{2},\beta_{3}$ from the tables presented in Ref.~\cite{Moller2016}. We present the results in terms of the $CP$-violating $\pi$-meson - nucleon interaction constants $\bar{g}_{0},\bar{g}_{1}$ and $\bar{g}_{2}$, the QCD parameter $\bar{\theta}$ and the quark chromo-EDMs. In Section \ref{EnhancementComparisonSection}, we quantitatively assess the enhancement provided from the octupole/quadrupole mechanisms relative to the current material of choice in the CASPEr electric experiment. In Section \ref{SolidStateSection}, we discuss the implications of the enhancement mechanisms for solid state experiments which may benefit from enhanced nuclear moments in their search for axion dark matter, and subsequently make suggestions for potential alternative candidate solid state compounds. The application of the enhancement mechanisms to experiments aiming to detect the so-called piezoaxionic effect is also discussed, with a list of potential candidate crystals presented. 

% ------------------ Original final paragraph of introduction ----------------------

% In this paper we perform estimates of the nuclear Schiff moment in various stable and long-living nuclei theorised to have either a static octupole deformation or a dynamical octupole vibration mode, using values of $\beta_{2},\beta_{3}$ from the tables presented in Ref.~\cite{Moller2016}. We present the results in terms of the $CP$-violating $\pi$-meson - nucleon interaction constants $\bar{g}_{0},\bar{g}_{1}$ and $\bar{g}_{2}$, the QCD parameter $\bar{\theta}$ and the quark chromo-EDMs. In Section \ref{SolidStateSection}, we discuss the implications of the octupole mechanism for solid state experiments which may benefit from an enhanced Schiff moment in their search for axion dark matter. We quantitatively assess the enhancement provided from this mechanism relative to the current material of choice in the CASPEr electric experiment, and subsequently make suggestions for potential alternative candidate solid state compounds. The application of the octupole mechanism to experiments aiming to detect the so-called piezoaxionic effect is also discussed, with a list of potential candidate crystals presented. We also discuss effects produced by the nuclear MQM.

% ---------------------------------------------------------------------------------

\section{Estimate of the enhanced nuclear Schiff Moment in nuclei with a theorised octupole deformation} \label{EstimatesSection}
In this section, we will present our updated estimates of the nuclear Schiff moments in $^{153}$Eu, $^{153}$Sm, $^{155}$Gd, $^{161}$Dy, $^{163}$Dy, $^{165}$Er, $^{221}$Fr, $^{223}$Fr, $^{223}$Rn $^{225}$Ac, $^{225}$Ra, $^{227}$Ac, $^{229}$Th, $^{229}$Pa, $^{233}$U, $^{235}$U, $^{237}$Np and $^{239}$Pu assuming that these nuclei have static or dynamical octupole deformation. These results are presented in terms of the CP-violating $\pi$-meson - nucleon interaction constants $\bar{g}_{0},\bar{g}_{1}$ and $\bar{g}_{2}$, the QCD parameter $\bar{\theta}$ and the quark chromo-EDMs. We note that the first indication that a nucleus may  have an octupole deformation comes via an analysis of the rotational spectra, which appears to resemble the spectra of a diatomic molecule of different atoms~\footnote{The enhancement of the EDM and Schiff moment in nuclei with an octupole deformation is similar to the enhancement of the $T$,$P$-violating effects in polar molecules with non-zero electron angular momentum which have doublets of opposing parity.~\cite{Sushkov1978}. The doublet splitting in molecules is due to the Coriolis interaction. In nuclei, the splitting is dominated by the ”tunnelling” of the octupole bump to other side of the nucleus causing a change
of the valence nucleon spin projection to the nuclear axis. In fact, it is just an octupole vibration mode, so there is no sharp boundary between the static deformation in the minimum of the
potential energy and a soft octupole vibration when this minimum is very shallow or does not exist. Note that contrary to the Coriolis splitting in diatomic molecules the doublet splitting due the tunnelling does not increase with rotational angular momentum - see the nuclear spectra in Ref.~\cite{nudat3}.} 

The Schiff moment is defined by the following expression~\cite{Sushkov1984}

\begin{align}
   { \bf S} = \frac{e}{10} \left[ \ev{r^{2} {\bf r}} - \frac{5}{3Z} \ev{r^{2}} \ev{ {\bf r}} \right]\,,
\end{align}
where $e$ is the elementary charge and $Z$ is the atomic number. Here, $\ev{r^{n}} \equiv \int \rho({\bf r}) r^{n} d^{3}r$ represents the moments of the nuclear charge density $\rho$. If a nucleus has an octupole deformation $\beta_{3}$ and a quadrupole deformation $\beta_{2}$, the Schiff moment in the fixed body frame is proportional to the octupole moment $O_{\rm{intr}}$~\cite{Auerbach1996,Spevak1997}

\begin{align} \label{Sintri}
    S_{\rm{intr}} \approx \frac{3}{5 \sqrt{35}} O_{\rm{intr}} \beta_{2} \approx \frac{3}{20 \pi \sqrt{35}} e Z R^{3} \beta_{2} \beta_{3} \,,
\end{align}
where $R$ is the nuclear radius. However both Schiff and electric dipole moments are forbidden in the laboratory frame, due to parity and time reversal invariance.

Nuclei with an octupole deformation and non-zero nucleon angular momentum have a doublet of nearby rotational states, with opposing parity $\ket{I^{\pm}}$ and the same angular momentum $I$ 
\begin{align} \label{doublet}
\ket{ I^{\pm} }=\frac{1}{\sqrt{2}} (\ket{\Omega} \pm \ket{-\Omega})\,, 
\end{align}
where $\Omega=\Sigma +\Lambda$ is  the projection of $I$ on to the nuclear axis. The states of this doublet are mixed by the $T$,$P$-violating interaction $W$ with mixing coefficient

\begin{align} \label{mixing}
    \alpha_{\pm} = \frac{\mel{I^{-}}{W}{I^{+}}}{E_{+} - E_{-}} \,.
\end{align}
We may express the nuclear $T$,$P$-violating potential $W$ as
\begin{align} \label{TPviolatingpotential}
    W = \frac{G}{\sqrt{2}} \frac{\eta}{2m} (\sigma \nabla) \rho \,,
\end{align}
where $G$ is the Fermi constant, $m$ is the nucleon mass, $\rho$ is the nuclear number density and $\eta$ is a dimensionless strength constant. This mixing polarises the nuclear axis ${ \bf n}$ along the axis of nuclear spin ${ \bf I} $

\begin{align}
    \ev{n_{z}} = 2 \alpha_{\pm} \frac{I_{z}}{I+1} \,,
\end{align}
and the intrinsic Schiff moment shows up in the laboratory frame~\cite{Auerbach1996,Spevak1997}

\begin{align} \label{Slab}
    S = 2 \alpha_{\pm} \frac{I}{I+1} S_{\rm{intr}} \,.
\end{align}
In accordance with Ref.~\cite{Spevak1997}, in nuclei with an octupole deformation, the $T$,$P$-violating matrix element may be approximated as

\begin{align} \label{melapprox}
    \mel{I^{-}}{W}{I^{+}} \approx \frac{\beta_{3} \eta}{A^{1/3}} [ \rm{eV}] \,,
\end{align}
given the number of nucleons $A$. Therefore, combining equations (\ref{Sintri}), (\ref{doublet}), (\ref{mixing}) and (\ref{Slab}), we may now write down an analytical estimate for the Schiff moment~\cite{Spevak1997,Feldmeier2020}

\begin{align} \label{SchiffMoment}
    S \approx 1 \cdot 10^{-4} \frac{I}{I+1} \beta_{2} (\beta_{3})^{2} Z A^{2/3} \frac{[\rm{KeV}]}{E_{+} - E_{-}} e  \eta \ [\rm{fm}^{3}] \,.
\end{align}
This expression is in agreement with numerical calculations available for a number of nuclei~\cite{Spevak1997}. 

Within meson exchange theory, the $\pi$-meson exchange gives the dominating contribution to the $T$,$P$-violating nuclear forces~\cite{Sushkov1984}. Using standard notation,
$g$ is the strong $\pi$-meson - nucleon interaction constant
and $\bar{g}_{0}$, $\bar{g}_{1}$, $\bar{g}_{2}$ are the $\pi$-meson - nucleon $CP$-violating interaction constants in the isotopic channels $T$ = 0, 1, 2.
%~\footnote{We have also estimated the contribution of the exchange by $\eta$-meson which is 4 times heavier than $\pi$-meson and is usually assumed to give a smaller contribution. Indeed, the second power of the meson mass appears in the denominator of the effective interaction constant for the meson-induced nucleon interaction, so the expected suppression is 1/16. However, the $\eta$-meson $CP$-violating exchange constant $\bar{g}$ is an order of magnitude larger than the $\pi$-meson constant $\bar{g}_{0}$~\cite{DeVries2015}. In addition, the $\eta$-meson $CP$-violating contribution has the same sign for protons and neutrons, which is contrary to the $\pi$-meson contribution. As a result, the suppression of the
%$\eta$-meson contribution to the nuclear Schiff moment is not as large as expected. Future more accurate calculations of the Schiff moment should include the $\eta$-meson contribution as well as the finite nuclear size corrections found in Ref.~\cite{Ginges2002}}.

Given these, one can express the results for the Schiff moment in terms of more fundamental parameters such as the QCD $\theta$-term constant
$\bar{\theta}$ using the relation $| g \bar{g}_{0}| $ = 0.21 $|\bar{\theta}|$ and $| g \bar{g}_{1}| $ = -0.046 $|\bar{\theta}|$~\cite{Demille2014,Bsaisou2015,Yamanaka2017}. Alternatively, the results can be expressed via the quark
chromo-EDMs $\tilde{d}_{u}$ and $\tilde{d}_{d}$: $g \bar{g}_{0}$ = $0.8 \cdot 10^{15}(
\tilde{d}_{u} + \tilde{d}_{d})$/cm, $g \bar{g}_{1}$ = $4 \cdot 10^{15}( \tilde{d}_{u} - \tilde{d}_{d})$/cm~\cite{PR}.

Numerical calculations performed in Ref.~\cite{Engel2003} found the Schiff moment of $^{225}$Ra, expressing it in terms of $g \bar{g}_{0,1,2}$

\begin{align} \label{SchiffRa1}
    S(^{225}\text{Ra},g) \approx (-2.6g \bar{g}_{0} + 12.9 g \bar{g}_{1} - 6.9 g \bar{g}_{2}) \ e \cdot \text{fm}^{3} \,. 
\end{align}
This expression may be rewritten to express the Schiff moment in terms of $\bar{\theta}$ and $\tilde{d}_{u}, \tilde{d}_{d}$

\begin{align}
\begin{split} \label{SchiffRa2}
    S(^{225}\text{Ra},\bar{\theta}) & \approx - \bar{\theta} \ e \cdot \text{fm}^{3} \,, \\
    S(^{225}\text{Ra},\tilde{d}) & \approx 10^{4} ( 0.50 \tilde{d}_{u} - 0.54 \tilde{d}_{d}) \ e \cdot \text{fm}^{2} \,.
\end{split}
\end{align}

The analytical expression for Eq. (\ref{SchiffMoment}) describes the dependence of the Schiff moment on various nuclear parameters. Following Refs.~\cite{Flambaum2020,Feldmeier2020,Dzuba2020}, we may use Eqs. (\ref{SchiffMoment},\ref{SchiffRa1},\ref{SchiffRa2}) to express the Schiff moment of all nuclei with an octupole deformation in the following form

\begin{align}
    S(g) & \approx K_{S} (-2.6g \bar{g}_{0} + 12.9 g \bar{g}_{1} - 6.9 g \bar{g}_{2}) e \cdot \text{fm}^{3} \,, \\
    S(\bar{\theta}) & \approx - K_{S} \bar{\theta} \ e \cdot \text{fm}^{3} \,, \label{SchiffThetaRa} \\
    S(\tilde{d}) & \approx 10^{4} K_{S} ( 0.50 \tilde{d}_{u} - 0.54 \tilde{d}_{d}) \ e \cdot \text{fm}^{2} \,,
\end{align}
where $K_{S} = K_{I} K_{\beta} K_{A} K_{E}$, given the following definitions

\begin{align}
\begin{split}
    K_{I} & = \frac{3I}{I+1} \,, \\
    K_{A} & = 0.00031 Z A^{2/3} \,,
\end{split}
\begin{split}
    K_{\beta} & = 791 \beta_{2} (\beta_{3})^{2} \,, \\
    K_{E} & = \frac{55 \text{KeV}}{E_{+} - E_{-}} \,.
\end{split}
\end{align}
%The numerical factors were chosen such that these parameters equal unity for $^{225}$Ra, and are of the order of unity for other heavy nuclei with an octupole deformation.
The numerical factors were chosen such that they are of the order of unity for heavy nuclei with an octupole deformation. Using these expressions we have calculated the value of $K_{S}$ for a select number of candidate nuclei using the tabulation of the parameters $\beta_{2}$ and $\beta_{3}$, which can be found in Ref.~\cite{Moller2016} for the ground states of nearly all nuclei. We note that estimates with other nuclear structure models can give different results~\cite{Ebata2017,Nazarewicz1984}. In nuclei which do not have static octupole deformation according to Ref.~\cite{Moller2016}, but still exhibit a significant dynamical octupole deformation, we have estimated the value of the squared octupole deformation parameter $\beta_{3}^{2}$ using the collective $B(E3)$ octupole transition probability for neighbouring even-even nuclei found in e.g. Ref.~\cite{Kibedi2002}. This transition probability is related to $\beta_{3}$ via the expression

\begin{align}
    B(E3)_{0^{+} \rightarrow 3} = \left( \frac{3}{4 \pi} \right)^{2} (Z e R_{0}^{3})^{2} \ev{\beta_{3}^{2}} \,.
\end{align}
The energy splittings of the octupole doublet $\Delta E_{\pm} = E_{+} - E_{-}$ were found in the database in Ref.~\cite{nudat3}. Thus, our estimates for the parameter $K_{S}$ are presented in Table \ref{TableKs}.

\begin{table}[!h]
\begin{center}
\setlength{\tabcolsep}{26pt}
\begin{tabular}{ccc}
\hline \hline
    $Z$  &  Isotope  & $K_{S}$  \\  \hline
    63 &$^{153}$Eu &0.99\\
    62 &$^{153}$Sm &2.4\\
    64 &$^{155}$Gd &1.2\\
    66 &$^{161}$Dy &4.3\\
    66 &$^{163}$Dy &0.31\\
    68 &$^{165}$Er &1.8\\
    87 &$^{221}$Fr &0.85\\
    87 &$^{223}$Fr &1.2\\
    86 &$^{223}$Rn &1.7\\
    89 &$^{225}$Ac &5.5\\
    88 &$^{225}$Ra &2.4\\
    89 &$^{227}$Ac &6.0\\
    90 &$^{229}$Th &1.2\\
    91 &$^{229}$Pa & 560 (2040)~\footnote{The estimate in parenthesis is presented with the caveat that the existence of the very close nuclear energy doublet $\Delta E_{\pm} \approx 60$ eV in $^{229}$Pa is confirmed.} \\
    92 &$^{233}$U &0.44\\
    92 &$^{235}$U &1.9\\
    93 &$^{237}$Np &2.6\\
    94 &$^{239}$Pu &0.2\\
 \hline \hline
\end{tabular}
\end{center}
\caption{Estimates of collective Schiff moments for octupole deformed nuclei which contain a low lying energy doublet to the ground state with identical spin and opposing parity.}
\label{TableKs}
\end{table}

The values of $K_{S}$ presented in Table \ref{TableKs} may be directly compared to those calculated in Ref.~\cite{Dzuba2020}. With the exception of $^{229}$Pa~\footnote{Which is several orders of magnitude larger when assuming a ground state energy splitting of $\Delta_{\pm} E \approx 60$ eV}, the values of $K_{S}$ for $^{153}$Eu, $^{161}$Dy, $^{163}$Dy, $^{221}$Fr, $^{223}$Fr, $^{223}$Rn, $^{225}$Ac, $^{227}$Ac, $^{229}$Th, $^{233}$U, $^{235}$U, $^{237}$Np and $^{239}$Pu are of a similar order to the estimates made in Ref.~\cite{Dzuba2020}, in which values for the octupole deformation parameter $\beta_{3}$ were taken from a range of sources. Further, this method allows for the slight improvement on the estimates made in Ref.~\cite{Dzuba2020} for the nuclei $^{161}$Dy, $^{163}$Dy, $^{229}$Th, $^{233}$U, $^{235}$U and $^{239}$Pu, all which are theorised to have a soft dynamical octupole vibration mode. Whilst we note that the existence of such a small energy splitting in $^{229}$Pa is yet to be confirmed, the various theoretical results which claim the existence of both a static and dynamical octupole deformation in each of the nuclei presented in Table \ref{TableKs} must be confirmed via experimental probing of the nuclear rotational spectra.

% \section{Implications of the enhancement of nuclear moments for solid state experiments searching for axion dark matter}
\section{Effects of enhanced nuclear moments in deformed nuclei}
\label{EnhancementComparisonSection}

\subsection*{Enhanced nuclear Schiff moments in octupole deformed nuclei}

As aforementioned, there are various detection techniques which may be used to detect $T$ and $P$-violating atomic, molecular and neutron EDMs. Specifically, experiments involving Pb based solid state samples are the current area of interest, see Refs.~\cite{Mukhamedjanov2005, Bouchard2008}. Further, experiments searching for ultralight dark matter such as ones conducted by the CASPEr collaboration measure the axion-induced Pb Schiff moment in solid state materials. 

It may however be interesting to consider performing such experiments with solid state samples containing atoms considered in Section \ref{EstimatesSection}, as they may provide an  advantage, due to the enhancement of the Schiff moment via the octupole mechanism. This may be quantitatively assessed via a comparison of the coefficients $K_{S}$ to that of $^{207}$Pb. 

$^{207}$Pb is considered to be nearly spherical. It has identical spin and parity to $^{199}$Hg ($I^{P} = 1/2^{-}$) and a similar nuclear magnetic moment: $\mu_{^{207}\rm{Pb}} = 0.59 \mu_{N}$, $\mu_{^{199}\rm{Hg}} = 0.51 \mu_{N}$, where $\mu_{N}$ is the nuclear magneton. The Schiff moment of $^{199}$Hg was first calculated in Ref.~\cite{Sushkov1986}, with more complete calculations in terms of the $\pi$-meson interaction constants $g \bar{g}_{0,1,2}$ using 5 different interaction models presented in Ref.~\cite{Ban2010}. 

Using the above arguments, we may consider the Schiff moments of $^{199}$Hg and $^{207}$Pb to be approximately equal. Taking the average of the results presented in Ref.~\cite{Ban2010}, and once again expressing the results in terms of $\bar{\theta}$ and the quark chromo-EDMs $\tilde{d}_{u}$ and $\tilde{d}_{d}$ we have

\begin{align}
    S(^{207} \text{Pb},g) & \approx S(^{199} \text{Hg},g) \\ \nonumber
    & \approx (0.023 g \bar{g}_{0} -0.007 g \bar{g}_{1} +0.029  \bar{g}_{2}) \ e \cdot \text{fm}^{3} \,, \\
    S(^{207} \text{Pb},\theta) &  \approx S(^{199} \text{Hg},\theta)  \approx 0.005 \ \bar{\theta} \ e \cdot \text{fm}^{3} \,, \label{SchiffThetaPb} \\
    S(^{207}\text{Pb},\tilde{d}) &  \approx S(^{199} \text{Hg},\tilde{d})  \approx 5 \tilde{d}_{d} \ e \cdot \text{fm}^{2} \,.
\end{align}
We may now assess the relative enhancement of the estimated nuclear Schiff moments calculated in Section \ref{EstimatesSection} to that of $^{207}$Pb by defining the parameter $K_{S,\text{Pb}}$ to be the ratio of Eqs. (\ref{SchiffThetaRa}) and (\ref{SchiffThetaPb}). We can further estimate the Schiff moment induced energy shift in compounds containing these nuclei via the scaling $ \delta E \sim  Z^{2} R_{S} K_{S,\text{Pb}} $, where $R_{S} = (R_{S,1/2} + 2R_{S,3/2})/3$ is the average relativistic factor, which is different for $p_{1/2}$ and $p_{3/2}$ electrons~\cite{Sushkov1984}

\begin{table}[ht]
\centering
\setlength{\tabcolsep}{9pt}
\begin{tabular}{cccccc}
\hline
\hline 
    $Z$  &  Isotope & $I$ & $K_{S,\text{Pb}}$  &  $K_{Z,\text{Pb}}$ & $K_{\text{Tot}}$  \\  \hline
    63 &$^{153}$Eu & 5/2 & 199 & 0.310 & 61.6\\
    62 &$^{153}$Sm & 3/2 & 480 & 0.291 & 140 \\
    64 &$^{155}$Gd & 3/2 & 235 & 0.330 & 77.5 \\
    66 &$^{161}$Dy & 5/2 & 852 & 0.373 & 318 \\
    66 &$^{163}$Dy & 5/2 & 62.6 & 0.373 & 23.3 \\
    68 &$^{165}$Er & 5/2 & 350 & 0.422  & 148 \\
    87 &$^{221}$Fr & 5/2 & 171 & 1.37 & 234 \\
    87 &$^{223}$Fr & 3/2 & 248 & 1.37 & 339 \\
    86 &$^{223}$Rn & 7/2 & 348 & 1.28 & 446 \\
    89 &$^{225}$Ac & 3/2 & 1100 & 1.56 & 1710 \\
    88 &$^{225}$Ra & 1/2 & 473 & 1.46 & 690 \\
    89 &$^{227}$Ac & 3/2 & 1210 & 1.55 & 1880 \\
    90 &$^{229}$Th & 5/2 & 238 & 1.66  & 395 \\
    91 &$^{229}$Pa & 5/2 & 111000 & 1.77 & 197000 \\
    92 &$^{233}$U  & 5/2 & 88.7 & 1.89  & 167 \\
    92 &$^{235}$U  & 7/2 & 376 & 1.88 & 708 \\
    93 &$^{237}$Np & 5/2 & 525 & 2.01 & 1060 \\
    94 &$^{239}$Pu & 1/2 & 32.0 & 2.14 & 68.6 \\
 \hline \hline
\end{tabular}
% \caption{Ratios of atomic factors to those for $^{207}$Pb in nuclei with an octupole deformation.}
\caption{Ratios of the nuclear Schiff moment induced energy shift in solids containing nuclei with an octupole deformation to those containing $^{207}$Pb.}
\label{KCoefficientTable}
\end{table}

\begin{align}
    R_{S,1/2} & \approx \frac{4 \gamma_{1/2} x_{0}^{2 \gamma_{1/2} -2}}{[\Gamma (2 \gamma_{1/2} +1)]^{2}} \,, \\
    R_{S,3/2} & \approx \frac{48 \gamma_{1/2} x_{0}^{ \gamma_{1/2} + \gamma_{3/2} -3}}{\Gamma (2 \gamma_{1/2} +1)\Gamma (2 \gamma_{3/2} +1)} \,.
\end{align}
Here, $\gamma_{1/2} = \sqrt{1 - Z^{2} \alpha^{2}}$; $\gamma_{3/2} = \sqrt{4 - Z^{2} \alpha^{2}}$, $\alpha$ is the fine structure constant, $\Gamma(x)$ is the standard $\Gamma$ function and $x_{0} = 2Z R_{0}/a_{B}$, where $R_{0} = 1.2 \ [\text{fm}] \ A^{1/3}$ is the nuclear radius and $a_{B}$ is the Bohr radius. Our estimates of these atomic factors are listed in Table \ref{KCoefficientTable}, where we have further defined the coefficients

\begin{align}
\begin{split}
    K_{Z,\text{Pb}} = \frac{Z^{2} R_{S}}{Z_{\text{Pb}}^{2} R_{S,\text{Pb}}} \,,
\end{split}
\begin{split}
K_{\text{Tot}} = K_{S,\text{Pb}} \cdot K_{Z,\text{Pb}} \,.
\end{split}
\end{align}

The estimates presented in Table \ref{KCoefficientTable}, as expected, indicate a substantial enhancement of the relevant atomic factors relative to $^{207}$Pb in nuclei which are theorised to have an octupole deformation. Given this, one may consider the use of compounds containing these nuclei in various experiments which aim to measure the energy shift induced by the nuclear Schiff moment in solid state compounds. 

\subsection*{Enhanced nuclear magnetic quadrupole moments in quadrupole deformed nuclei}

In a similar way, we may also assess the potential advantage of performing measurements in quadrupole deformed nuclei. Contrary to the case of octupole deformation, quadrupole deformation is a well studied nuclear property which is indicated, for example, by collective electric quadrupole moments which may exceed single-particle quadrupoles by up to two orders of magnitude. We will begin by providing a brief overview of the MQM  enhancement mechanism in such nuclei, following~\cite{Lackenby2018}. The magnetic quadrupole moment of a nucleus due to the electromagnetic current of a single nucleon with mass $m$ is defined by the second order tensor operator~\cite{Khriplovich1991}

\begin{align} \label{MQMExpression}
\begin{split}
    \hat{M}^{\nu}_{kn} = \frac{e}{2m} \biggl[  3 \mu_{\nu} \biggl( r_{k} \sigma_{n} + \sigma_{k} r_{n} - \frac{2}{3} \delta_{kn} \boldsymbol{\hat{\sigma}} { \bf r} \biggr) \\ + 2 q_{\nu} (r_{k} l_{n} + l_{k} r_{n})  \biggr] \,,
\end{split}
\end{align}
where $\nu = p,n$ for protons and neutrons respectively and $\mu_{\nu}, q_{\nu}$ are the magnetic moment and charge of the nucleon respectively. As the nuclear MQM is both $T$ and $P$ violating, it is forbidden in the absence of nucleon EDMs and $T$,$P$-violating nuclear forces. It is understood that the open shell nucleons interact with the core of the nucleus via the $T$,$P$-violating potential (\ref{TPviolatingpotential})~\cite{Khriplovich1991,Sushkov1984,Flambaum1994}, resulting in a perturbed ``spin-hedgehog'' wave function of a nucleon given by~\cite{Khriplovich1991,Flambaum1994}

\begin{align} \label{spinhedgehog}
    \ket{\psi^{\prime}} & = \left( 1 + \frac{\xi_{\nu}}{e} \boldsymbol{\hat{\sigma}} \boldsymbol{\hat{\nabla}} \right) \ket{\psi_{0}} \,, \\
    \xi_{n} & \approx -2 \times 10^{-21} \eta_{\nu} \ e \cdot cm \,,
\end{align}
where once again $\nu = p,n$ for protons and nucleons respectively and $\eta_{\nu}$ represent the $T$,$P$-violating nuclear strength constants in the $T$,$P$-violating potential (\ref{TPviolatingpotential}). This spin-hedgehog contributes to the $T$,$P$-violating effects in nuclei, which results in an enhanced collective MQM in deformed nuclei~\cite{Flambaum1994}. Using Eqs. (\ref{MQMExpression}) and (\ref{spinhedgehog}), the MQM for a single nucleon due to the $T$,$P$-violating valence-core interaction is given by 

\begin{align}
    M^{TP} = M^{TP}_{zz} = \xi \frac{2}{m} ( \mu \ev{\sigma \cdot l} - q \ev{\sigma_{z} l_{z}}) \,.
\end{align}
In the Nilsson basis~\cite{Nilsson1955} the nucleon's total angular momentum projection onto the symmetry axis is given by $\Omega_{N} = \Lambda_{N} + \Sigma_{N}$, where $\Sigma_{N} = \pm 1/2$ is the spin projection and $\Lambda$ is the orbital angular momentum projection of the nucleon. In this basis, the MQM generated by the spin-hedgehog in Eq. (\ref{spinhedgehog}) is

\begin{align}
    M_{\nu}^{TP} = 4 \Sigma_{N} \Lambda_{N} \xi (\mu_{\nu} - q_{\nu}) \frac{\hbar}{m_{p}c} \,.
\end{align}
Further, the  permanent nucleon EDM $d_{\nu}$ also contributes to the nuclear MQM
%, with proportionality $d_{\nu} \propto M_{\nu}^{EDM}$
~\cite{Khriplovich1976}. The result of the calculation in the Nilsson model is ~\cite{Lackenby2018}
%As both protons and neutrons are expected to have an EDM, both will provide contributions to the MQM. 
%Using a valence nucleon approach, the ratio of the two contributions $M_{\nu}^{TP}/M_{\nu}^{EDM}$ is independent of the total angular momentum of the nucleon $I$:% Thus, up to nondiagonal elements of the definite $I$, the ratio is the same in the Nilsson model:

\begin{align}
    M_{\nu}^{EDM} \approx 4 \Sigma_{N} \Lambda_{N} d_{\nu} \frac{\hbar}{m_{p} c} \,.
\end{align}
Thus, the MQM generated by a single nucleon is given by

\begin{align}
    M_{\nu} = 4 \Sigma_{N} \Lambda_{N} M_{\nu}^{0} \,,
\end{align}
where
\begin{align}
    M_{\nu}^{0} = [ \xi (\mu_{\nu} - q_{\nu}) d_{\nu}] \frac{\hbar}{m_{p} c} \,.
\end{align}
Specific values of the nuclear MQM in this mechanism were calculated in quadrupole deformed nuclei of experimental interest in Ref.~\cite{Lackenby2018}, by summing up the contributions from each nucleon. 

An enhanced MQM in a nucleus also results in an enhanced MQM induced energy shift in solids containing this nucleus. Thus, in a similar way to the previous section, we may estimate this expected enhancement relative to the Schiff moment induced energy shift in Pb based solids. Using Ref.~\cite{Sushkov1984}, the ratio of the MQM and Schiff moment contributions to atomic EDM may be approximated to be

\begin{align} \label{MQMratioanalytical}
    \frac{\delta E_{M}}{\delta E_{S}} \approx \frac{1}{12} \frac{M R_{M} \alpha a_{B}}{S R_{S}} \,.
\end{align}
Here $M$ is the magnetic quadrupole moment. The nuclear charge $Z$ dependence of the MQM and Schiff moment atomic matrix elements is similar ($\propto Z^2$) except for the relativistic factors. For the MQM, the relativistic factor is 

\begin{align}
    % x  & = \frac{\zeta_{1} \zeta_{2}}{| \zeta_{1} \zeta_{2}|} \frac{96 (\zeta_{1} + \zeta_{2} -2)(j_{1} + j_{2} -2)!}{(j_{2} - j_{1} + 2)! (j_{1} + j_{2} +3 )!(j_{1} - j_{2} +2)!} \,, \\
    % R_{M} & = \frac{(j_{2} - j_{1} + 2)!(j_{1} + j_{2} +3 )!(j_{1} - j_{2} +2)! \Gamma(\gamma_{1} + \gamma_{2}-2)}{(j_{1} + j_{2} -2)!\Gamma(\gamma_{2} - \gamma_{1}+3)\Gamma(\gamma_{1} + \gamma_{2}+3)\Gamma(\gamma_{1} - \gamma_{2}+3)} \,,
    R_{M} = 720 \frac{ \Gamma(\gamma_{j_{1}} + \gamma_{j_{2}}-2)}{\Gamma(\gamma_{j_{2}} - \gamma_{j_{1}}+3)\Gamma(\gamma_{j_{1}} + \gamma_{j_{2}}+3)\Gamma(\gamma_{j_{1}} - \gamma_{j_{2}}+3)} \,, 
\end{align}
where once again $\gamma_{j} = \sqrt{\left( j + 1/2 \right)^{2} - Z^{2} \alpha^{2}}$. This factor is formulated such that $R_{M} \rightarrow 1 $ as $Z \alpha \rightarrow 0$. 
%In the numerical estimates we use Schiff moment and nuclear charge of $^{207}$Pb.

% The factor of $1/2$ in Eq. (\ref{MQMratioanalytical}) accounts for the the matrix element $s \rightarrow p_{1/2}$ which exists for Schiff moments, but does not exist for MQMs.   
% where $j= l + s$ is the total angular momentum of the electron and
% \begin{align}
%     \zeta_{i} & = (-1)^{j_{i}+(1/2)-l_{i}}\left( j_{i} + \frac{1}{2} \right) \,.
% \end{align}

The proton and neutron MQM contributions can be expressed via the QCD $CP$-violating parameter $\bar{\theta}$ as~\cite{Lackenby2018}

\begin{align}
\begin{split}
    M_{p}^{0} (\bar{\theta}) & = 1.9 \times 10^{-3} \bar{\theta} \ e \cdot \text{fm}^{2} \,, \\
    M_{n}^{0} (\bar{\theta}) & = 2.5 \times 10^{-3} \bar{\theta} \ e \cdot \text{fm}^{2} \,.
\end{split}
\end{align}
Thus, using Eq. (\ref{MQMratioanalytical}), we have estimated the enhancement of the MQM induced energy shift relative to the energy shift due to the Schiff moment in $^{207}$Pb, see Table \ref{MQMTable}. We note that there appears to be a substantial enhancement of the nuclear MQM induced effects in systems containing quadrupole deformed nuclei when compared to the $^{207}$Pb nuclear Schiff moment induced effects. 

The enhancement provided by the quadrupole (MQM induced) mechanism, on average, seems to be of a similar order to the enhancement provided by the octupole (Schiff moment induced) mechanism. However, there is an important difference between these mechanisms. The Schiff moment induced mechanism for enhancement currently relies on the confirmation of the existence of static/dynamical octupole deformation in the nuclei of interest. This is not the case for the MQM induced mechanism, where  the existence of nuclear quadrupole deformation is firmly established through the measurement of the electric quadrupole moment in a wide range of nuclei, see Ref.~\cite{StoneQuad2016}. Therefore at this stage, the MQM mechanism for enhancement seems to be the more reliable when making considerations for experiment. Note however, that contrary to the Schiff moment effect, the MQM effect requires non-zero electron angular momentum and this leads to an increase of systematic effects, significantly limiting measurement accuracy.

\begin{table}[ht]
\centering
\setlength{\tabcolsep}{9pt}
\begin{tabular}{ccccc}
\hline
\hline 
    $Z$  &  Isotope & $I$ & $M$ & $|\delta E_M/\delta E_{S,\text{Pb}}|$  \\  \hline
    4 &$^{9}$Be & 3/2 & $0 M_{p}^{0} + 0.4 M_{n}^{0}$ & 1.2 \\
    10 &$^{21}$Ne & 3/2 & $0 M_{p}^{0} + 0.4 M_{n}^{0}$ & 1.2  \\
    13 &$^{27}$Al & 5/2 & $3 M_{p}^{0} + 4.5 M_{n}^{0}$ & 21  \\
    63 &$^{151}$Eu & 5/2 & $12 M_{p}^{0} + 23 M_{n}^{0}$ & 140  \\
    63 &$^{153}$Eu & 5/2 & $12 M_{p}^{0} + 20 M_{n}^{0}$ & 120  \\
    63 &$^{155}$Gd~\footnote{For an estimate, we use the result of $M$ from $^{153}$Eu in this isotope given the small difference in mass numbers.} & 3/2 & $12 M_{p}^{0} + 20 M_{n}^{0}$ & 124  \\
    63 &$^{157}$Gd~\footnote{Similarly, we use the result of $M$ from $^{153}$Eu in this isotope.} & 3/2 & $12 M_{p}^{0} + 20 M_{n}^{0}$ & 124  \\
    66 &$^{163}$Dy & 5/2 & $11 M_{p}^{0} + 21 M_{n}^{0}$ & 130  \\
    68 &$^{167}$Er & 7/2 & $21 M_{p}^{0} + 36 M_{n}^{0}$ & 240 \\
    70 &$^{173}$Yb & 5/2 & $14 M_{p}^{0} + 26 M_{n}^{0}$ & 170 \\
    72 &$^{177}$Hf & 7/2 & $17 M_{p}^{0} + 42 M_{n}^{0}$ & 270 \\
    72 &$^{179}$Hf & 9/2 & $20 M_{p}^{0} + 50 M_{n}^{0}$ & 230  \\
    73 &$^{181}$Ta & 7/2 & $19 M_{p}^{0} + 45 M_{n}^{0}$ & 290 \\
    90 &$^{229}$Th & 5/2 & $13 M_{p}^{0} + 27 M_{n}^{0}$ & 250  \\
    92 &$^{235}$U~\footnote{Similarly, we use the result of $M$ from $^{229}$Th in this isotope.} & 7/2 & $13 M_{p}^{0} + 27 M_{n}^{0}$ & 260  \\
 \hline \hline
\end{tabular}
\caption{Ratios of the nuclear MQM induced energy shift  to the energy shift produced by the Schiff moment in $^{207}$Pb. %Estimates have been performed basing on the ratio of the matrix elements between $s_{1/2} \rightarrow p_{3/2}$ states.
}
\label{MQMTable}
\end{table}

% \begin{table}[ht]
% \centering
% \setlength{\tabcolsep}{9pt}
% \begin{tabular}{ccccc}
% \hline
% \hline 
%     $Z$  &  Isotope & $I$ & $M$ & $|G_{\text{Pb}}|$  \\  \hline
%     4 &$^{9}$Be & 3/2 & $0 M_{p}^{0} + 0.4 M_{n}^{0}$ & 9.65 \\
%     10 &$^{21}$Ne & 3/2 & $0 M_{p}^{0} + 0.4 M_{n}^{0}$ & 9.72  \\
%     13 &$^{27}$Al & 5/2 & $3 M_{p}^{0} + 4.5 M_{n}^{0}$ & 166  \\
%     63 &$^{151}$Eu & 5/2 & $12 M_{p}^{0} + 23 M_{n}^{0}$ & 1100  \\
%     63 &$^{153}$Eu & 5/2 & $12 M_{p}^{0} + 20 M_{n}^{0}$ & 996  \\
%     66 &$^{163}$Dy & 5/2 & $11 M_{p}^{0} + 21 M_{n}^{0}$ & 1043 \\
%     68 &$^{167}$Er & 7/2 & $21 M_{p}^{0} + 36 M_{n}^{0}$ & 1900 \\
%     70 &$^{173}$Yb & 5/2 & $14 M_{p}^{0} + 26 M_{n}^{0}$ & 1380 \\
%     72 &$^{177}$Hf & 7/2 & $17 M_{p}^{0} + 42 M_{n}^{0}$ & 2120\\
%     72 &$^{179}$Hf & 9/2 & $20 M_{p}^{0} + 50 M_{n}^{0}$ & 2520  \\
%     73 &$^{181}$Ta & 7/2 & $19 M_{p}^{0} + 45 M_{n}^{0}$ & 2330 \\
%     90 &$^{229}$Th & 5/2 & $13 M_{p}^{0} + 27 M_{n}^{0}$ & 2000  \\
%     92 &$^{235}$U~\footnote{For an estimate, we use the result of $M$ from $^{229}$Th in this isotope given the small difference in mass numbers.} & 7/2 & $13 M_{p}^{0} + 27 M_{n}^{0}$ & 2100  \\
%  \hline \hline
% \end{tabular}
% \caption{Ratios of the nuclear MQM induced energy shift in nuclei with a quadrupole deformation to the energy shift produced by the Schiff moment in $^{207}$Pb. Estimates have been performed basing on the ratio of the matrix elements between $s_{1/2} \rightarrow p_{3/2}$ states.}
% \label{MQMTable}
% \end{table}

We also note that the atoms (or ions) presented in Table \ref{KCoefficientTable} which have nuclear spin $I\geq 1$ and a non-zero electron angular momentum also exhibit an enhanced nuclear MQM effect due to the octupole mechanism, see Ref.~\cite{Mansour2022}. However, on average, the contribution to the MQM from this mechanism is smaller than the contribution from the spin-hedgehog mechanism.

\section{Implications of enhanced nuclear moments for solid state experiments searching for axion dark matter} \label{SolidStateSection}

In this section, we discuss experimental applications of enhanced nuclear Schiff and magnetic quadrupole moments. In particular, the enhancement of these effects in deformed nuclei may lead to an improved sensitivity for solid state experiments searching for axion dark matter. 

\subsection*{Possible enhancement of CASPEr electric effect}

Here we present a few potential alternative candidate compounds for the CASPEr electric experiment. The oscillating axion dark matter background produces an oscillating nuclear Schiff moment. CASPEr electric aims to probe the effects of
this Schiff moment 
%interaction via a measurement of the resulting nuclear magnetic resonance 
in a polarised ferroelectric crystal.
%, in the presence of an external bias magnetic field.
Current measurements use $^{207}$Pb$^{2+}$ in a poled ferroelectric PMN-PT crystal (Lead Magnesium Niobate-Lead Titanate, chemical formula (PbMg$_{1/3}$Nb$_{2/3}$O$_{3}$)$_{2/3}$ - (PbTiO$_{3}$)$_{1/3}$)~\cite{CasperNew}. Let us consider the possibility of using another suitable ferroelectric candidate crystal which includes a stable isotope with an enhanced Schiff moment due to the octupole mechanism. 

% This interaction may be detected by magnetic resonance techniques. 

One possibility may be the replacing this compound with the ferroelectric Eu$_{1/2}$Ba$_{1/2}$TiO$_{3}$, which was proposed for use in the search for the electron EDM in Ref.~\cite{ASushkov2010}. The Eu$^{2+}$ ion has ground state $^{8}S_{7/2}$, with electron configuration [Xe]4f$^{7}$. This compound, being ferroelectric, has an extremely large effective electric field $E^{*} \approx 10 \ \text{MV/cm}$ in a poled crystal. As per our estimates presented in Table \ref{KCoefficientTable}, the Europium isotope exhibits a significantly enhanced Schiff moment compared to $^{207}$Pb, meaning one may also expect an enhanced Schiff moment-induced energy shift, which may be advantageous for the experiments performed by the CASPEr collaboration. The disadvantage for the Schiff moment measurement here is the non-zero electron angular momentum of Eu$^{2+}$. In this case, there are various systematic effects that arise from the use of such compounds, see for example the discussion in Ref.~\cite{ASushkov2010}. However, this compound is suitable for the MQM effect which requires non-zero electron angular momentum, similar to the electron EDM effect. It is interesting to note that the Eu$^{3+}$ ion has zero electron angular momentum in the ground state, so compounds containing this ion may be better for the measurement of the Schiff moment effect.  

% NOte that Eu 3+ may be of advantage as it has zero electron angular momentum in the ground state

Another potential candidate crystal is the ferroelectric-ferroelastic Gadolinium Molybdate (chemical formula Gd$_{2}$(MoO$_{4}$)$_{3}$), which was also mentioned in proposals for similar electron EDM experiments~\cite{ASushkov2009}. The Gd$^{3+}$ cation has the same electron configuration  and ground state as Eu$^{2+}$. As Gadolinium Molybdate is also a ferroelectric, a large effective electric field is produced in a poled crystal. This compound would similarly exhibit  enhanced MQM and Schiff moment-induced energy shifts compared to isotopes of Pb, making it another potential candidate compound for such experiments. 
% $E^{*} \approx 20 \ \text{MV/cm} $
%We note that one should not naively conclude that it would be advantageous to perform experiments using compounds which contain atoms exhibiting an enhanced nuclear Schiff moment due to the octupole mechanism or MQM due to the spin-hedgehog mechanism in deformed nuclei. However, the MQM effect requires non-zero electron angular momentum, similar to the electron EDM effect. 

\subsection*{Applications to experiments measuring the piezoaxionic effect}

Such enhanced nuclear moments may also prove to be advantageous in proposed measurements of the so-called \textit{piezoaxionic effect}, in which a $P$-violating axion-like dark matter background produces a stress in piezoelectric crystals~\cite{PiezoaxionicEffect}. Piezoelectric crystal structures violate parity. Therefore, no symmetry forbids the occurrence of stress upon application of an electric field, which is $P$-violating, but $T$-invariant. This is the converse piezoelectric effect~\cite{Lippmann1881,Curie1881}. Vice versa, an electric field results from an applied stress, i.e., the piezoelectric effect~\cite{PCurie1880,JCurie1880}. 

An axion-like dark matter produces oscillating $T$,$P$-violating moments~\cite{Graham2011,Stadnik2014}, which induce an atomic EDM creating an electric field. This field produces a stress across the piezoelectric crystal. Ref.~\cite{PiezoaxionicEffect} proposes the measurement of the bulk acoustic modes excited by this axion induced stress. It seems to be advantageous to select candidate piezoelectric crystals which contain one or more nuclei with an enhanced Schiff moment or MQM.

\begin{table}[!h]
\begin{center}
\setlength{\tabcolsep}{35pt}
\begin{tabular}{cc}
\hline \hline
   Class & Candidate \\ \hline
    \multirow{2}{*}{6} & $ {\rm {\bf Dy}_{3} Cu Ge Se_{7} }$ \\ 
   &  $ {\rm {\bf Dy}_{3} Cu Sn Se_{7}} $ \\[3mm] 

    32 & ${ \rm Na { \bf Dy} H_{2} S_{2} O_{9}}^{*}$ \\[3mm]

    6mm & $ {\rm {\bf Dy}_{3} Se_{4} O_{12}}$ \\[3mm] 
    
    \multirow{2}{*}{4mm} & ${ \rm {\bf Dy}  Si_{3} Ir }^{*}$ \\ 
    & ${ \rm {\bf Dy}  Ag Se_{2} }^{*}$ \\[3mm]
    
    $\bar{6}$m2 &  ${\rm {\bf Dy} Ta_{7} O_{19}}$ \\[3mm]

    \multirow{3}{*}{$\bar{4}3$m} & $ {\rm {\bf Dy} Ni Bi}$  \\
    & $ {\rm {\bf Dy}_{3} Sb_{4} Au_{3}}$  \\
    & $ {\rm {\bf Dy} Sb Pt}$  \\[1.5mm]
    
    mm2 & ${ \rm {\bf U}  C O_{5}}^{*} $ \\[3mm]
    
    \multirow{3}{*}{$\bar{4}2$m} & ${ \rm {\bf Dy}  Ag Te_{2} }^{*}$  \\
    & ${ \rm {\bf Dy}_{2} Be_{2} Ge O_{7} }^{*}$ \\
    & $ {\rm Ce_{2} {\bf Dy}_{2} O_{7}}$  \\[3mm]
    
    \multirow{3}{*}{3m} & $ {\rm {\bf Dy}Cu Se_{2}}$  \\
    & $ {\rm {\bf U } O F_{4}}^{*}$ \\
    & $ {\rm {\bf U } Cd}^{*}$ \\
    
    \hline \hline
\end{tabular}
\end{center}
\caption{Candidate crystals which may exhibit an enhanced energy shift due to the nuclear Schiff moment octupole mechanism, as found in Ref.~\cite{TheMaterialsProjectReference}. Nuclei which exhibit an enhanced Schiff moment due to the octupole mechanism are represented in bold. The compounds with asterisks were proposed in Ref.~\cite{PiezoaxionicEffect}.}
\label{PiezoelectricSampleTable}
\end{table}

Experiments aiming to detect the piezoaxionic effect require a candidate crystal which exhibits piezoelectricity. Further, it is required that the crystal is not ferromagnetic or strongly ferroelectric, in order to minimise any signal losses associated with movement of the domain walls. As such, the authors of Ref.~\cite{PiezoaxionicEffect} have proposed a list of candidate compounds containing either U or Dy for experiment~\footnote{These isotopes may also exhibit an enhanced nuclear MQM}. These crystals were chosen due to their structural similarity to well-developed piezoelectric materials already used in bulk resonators (see Table II in Ref.~\cite{PiezoaxionicEffect}). Using the the database of The Materials Project~\cite{TheMaterialsProjectReference} we have identified an additional number of crystals which may benefit from effects resulting from an enhanced Schiff moment and MQM. This database has calculated the piezoelectric tensor properties of a large number of materials, using Density Functional Perturbation Theory~\cite{Baroni1987,Baroni2001,Gonze1995}. These potential candidate crystals are listed in Table \ref{PiezoelectricSampleTable}. 

% The piezoelectric properties of these materials were predicted using density functional theory (DFT), and matched to an experimentally determined crystal structure.

% We once again note here the importance of experimentally confirming that the relevant isotopes of Dy and U do indeed exhibit a dynamical octupole vibration mode as theoretically predicted. 

\section{Summary}
Deformed nuclei possess enhanced $T$,$P$-violating  moments. Collective MQMs appear in nuclei with a quadrupole deformation, while nuclei with an octupole deformation have a collective electric octupole moment, EDM, Schiff moment and MQM in the intrinsic frame which rotates with the nucleus. It is theorised that nuclei in isotopes of Eu, Sm, Gd, Dy, Er, Fr, Rn, Ac, Ra, Th, Pa, U, Np and Pu contain either a static octupole deformation, or a soft dynamical octupole vibration mode. This implies that these nuclei have doublets of close opposite parity rotational states with the same spin, which are mixed by $T$,$P$-violating forces. This mechanism produces enhanced $T$,$P$-violating nuclear moments.

In Section \ref{EstimatesSection}, we focus on the enhanced nuclear Schiff moment due to this mechanism, presenting updated estimates for its value in isotopes of these nuclei using values of the static octupole deformation parameter $\beta_{3}$ presented in Ref.~\cite{Moller2016}. In nuclei which do not have static octupole deformation according to Ref.~\cite{Moller2016}, but still exhibit a significant dynamical octupole deformation, we have estimated the value of the squared octupole deformation parameter using the collective $B(E3)$ octupole transition probability for neighbouring even-even nuclei found in e.g. Ref.~\cite{Kibedi2002}. These estimates, presented in Table \ref{TableKs}, are of a similar order to the estimates presented in Ref.~\cite{Dzuba2020}, and are slightly improved in comparison to this reference for a range of nuclei which are theorised to exhibit a soft dynamical octupole vibration mode. 

In Section \ref{EnhancementComparisonSection}, we quantitatively assess the relative enhancement of the Schiff moment in atoms with nuclei which are theorised to exhibit an octupole deformation compared to $^{207}$Pb. The relative enhancement provided from the theorised octupole mechanism may be of an advantage to solid state experiments such as CASPEr electric, which is searching for axion dark matter via detection of the effects produced by an oscillating nuclear Schiff moment in a polarised ferroelectric crystal. We also assess the effects of an enhanced MQM due to the spin-hedgehog mechanism in quadrupole deformed nuclei. 

Section \ref{SolidStateSection} of this paper discusses the implications of an enhanced nuclear Schiff moment and magnetic quadrupole moment to the search for axion dark matter. We suggest a few potential alternative candidate solid state compounds containing atoms with  enhanced nuclear $P,T$-violating  moments for use in the CASPEr electric setup. In the last part of this section we briefly summarise the expected applications of the octupole/quadrupole mechanisms to proposed measurements of the so-called piezoaxionic effect, in which an axion dark matter background produces a stress across piezoelectric crystals. Building on the indicative list provided in the experimental proposal for the detection of this effect in Ref.~\cite{PiezoaxionicEffect}, we have included a number of further candidate crystals which may be used for measurements.  

% which has been collected from the database of The Materials Project~\cite{TheMaterialsProjectReference}.

We once again stress that prior to the commencement of any experiment using compounds in which the octupole mechanism is expected, the various theoretical results which claim both static and dynamical octupole deformation should first be confirmed via experimentally probing the nuclear rotational spectra. Whilst it is theorised that the nuclei presented in Table \ref{TableKs} exhibit either static or dynamical octupole deformation, the rotational spectra, which may be analysed using the database found in Ref.~\cite{nudat3}, in a large number of cases does not provide conclusive evidence. Note that there is no such problem for the enhanced nuclear magnetic quadrupole moments which do not require octupole nuclear  deformation.

\section{Acknowledgements}

This work was supported by the Australian Research Council Grants No.\ DP230101058 and DP200100150.

% \section{Estimates of Nuclear Schiff Moments}
% \begin{tabular}{ |p{3cm}p{3cm}p{2cm}| }
%  \hline
%  Z& Isotope & K{s}\\
%  \hline
%     64 &^{155}Gd &1.17\\
%     88 &^{225}Ra &2.3709\\
%     62 &^{153}Sm &2.40\\
%     68 &^{165}Er &1.75\\
%     63 &^{153}Eu &0.99\\
%     66 &^{161}Dy &4.26\\
%     66 &^{163}Dy &0.30\\
%     86 &^{223}Rn &1.65\\
%     87 &^{221}Fr &0.85\\
%     87 &^{223}Fr &1.47\\
%     89 &^{225}Ac &1.47\\
%     89 &^{227}Ac &5.51\\
%     90 &^{229}Th &6.04\\
%     91 &^{229}Pa & \\
%     92 &^{233}U &0.44\\
%     92 &^{235}U &1.88\\
%     93 &^{237}Np &2.36\\
%     94 &^{239}Pu &0.16\\
%  \hline
% \end{tabular}

\bibliographystyle{apsrev4-2}
\bibliography{References.bib}

\end{document}